\documentclass{ifacconf}

\usepackage{graphicx,xcolor}      
\usepackage{natbib}        
\usepackage{amsmath, amssymb}
\newtheorem{theorem}{Theorem}
\newtheorem{lemma}{Lemma}

\newtheorem{remark}{Remark}

\makeatletter
\DeclareRobustCommand{\qed}{%
	\ifmmode 
	\else \leavevmode\unskip\penalty9999 \hbox{}\nobreak\hfill
	\fi
	\quad\hbox{\qedsymbol}}
\newcommand{\openbox}{\leavevmode
	\hbox to.77778em{%
		\hfil\vrule
		\vbox to.675em{\hrule width.6em\vfil\hrule}%
		\vrule\hfil}}
\newcommand{\qedsymbol}{\openbox}
\newenvironment{proof}[1][\proofname]{\par
	\normalfont
	\topsep6\p@\@plus6\p@ \trivlist
	\item[\hskip\labelsep\itshape
	#1.]\ignorespaces
}{%
	\qed\endtrivlist
}
\newcommand{\proofname}{Proof}
\makeatother

\newenvironment{customproof}[1][Proof]
{\par\noindent\textit{#1}\ignorespaces}
{\hfill$\square$\par}

\renewcommand{\qed}{\hfill$\square$}

\begin{document}
\begin{frontmatter}

\title{Average Predictor-Feedback Control Design for Switched Linear Systems$^*$\thanksref{footnoteinfo}} 

\thanks[footnoteinfo]{Funded by the European Union (ERC, C-NORA, 101088147). Views and 
opinions expressed are however those of the authors only and do not necessarily reflect those 
of 	the European Union or the European Research Council Executive Agency. Neither the 
European Union nor the granting authority can be held responsible for them.}

\author[1]{Andreas Katsanikakis} 
\author[1]{Nikolaos Bekiaris-Liberis} 
\author[2]{Delphine Bresch-Pietri}

\address[1]{Department of Electrical and Computer Engineering, Technical University of Crete, Chania, 73100, Greece
}
\address[2]{MINES ParisTech, PSL Research University CAS-Centre Automatique et Systemes, 60 Boulevard Saint Michel 75006, Paris, France}

\begin{abstract}                
We develop an input delay-compensating feedback law for linear switched systems with time-dependent switching. Because the future values of the switching signal, which are needed for constructing an exact predictor-feedback law, may be unavailable at current time, the key design challenge is how to construct a proper predictor state. We resolve this challenge constructing an average predictor-based feedback law, which may be viewed as an exact predictor-feedback law for a particular average system without switching. We establish that, under the predictor-based control law introduced, the closed-loop system is exponentially stable, provided that the plant's parameters are sufficiently close to the corresponding parameters of the average system. In particular, the allowable difference is inversely proportional to the size of delay and proportional to the dwell time of the switching signal. Since no restriction is imposed on the size of delay or dwell time themselves, such a limitation on the parameters of each mode is inherent to the problem considered (in which no a priori information on the switching signal is available), and thus, it cannot be removed. The stability proof relies on two main ingredients—a Lyapunov functional constructed via backstepping and derivation of solutions' estimates for the difference between the average and the exact predictor states. We present consistent, numerical simulation results, which illustrate the necessity of employing the average predictor-based law for achieving stabilization and desired performance of the closed-loop system.
\end{abstract}


\end{frontmatter}

\section{Introduction} \label{sec:intro}

\setcounter{footnote}{1}
Switched systems appear in numerous applications including traffic flow control, \cite{traffic_flow}, automotive control, \cite{IOANNOY}, \cite{ACC}, networked control systems, \cite{telecom}, water networks, \cite{Water}, and epidemics spreading, \cite{SIR_D}. This type of systems may be also affected by the presence of delays in the input signal, which may degrade system's performance and stability when left uncompensated. For example, in vehicle dynamics, switching may appear due to changes between throttle/braking dynamics and input delays due to engine dynamics, \cite{IOANNOY}, \cite{ACC}. In fault diagnosis in water distribution systems, \cite{Water}, switching may appear in the modelling of actuator dynamics, (e.g., due to transition of a pump between off/on states), while input delays may appear due to water transport times. Consequently, it is essential to design control strategies that can handle simultaneously switching and input delay.

Related literature includes results on stability analysis of switched systems with input delay using approaches based on construction of Linear Matrix Inequalities (LMIs), \cite{Lin_discrete}, \cite{Xi_Corr}, \cite{Kp} and Lyapunov-Krasovskii functionals, \cite{Yue_Kras}, \cite{Mazenc_LKF}. All of these results impose a limitation on the size of delay or dwell time, which is not imposed here. Non-predictor-based control designs (typically restricting the delay size or dwell time) have been developed in \cite{Mazenc}, utilizing Lyapunov-based tools, and in \cite{Wu_Truncated}, \cite{Sakthi_Truncated} utilizing the truncated predictor-based method. In, \cite{Lin_LTI} a predictor-based controller is presented, which, however, requires availability of the future values of the switching signal for implementation. Our results can be also viewed as related to results dealing with hyperbolic Partial Differential Equation (PDE) systems with stochastic, \cite{Auriol}, or deterministic switching, \cite{Prieur}, as well as  systems with switched (stochastic or deterministic) input delays, \cite{Kong_P1}, \cite{Kong}, \cite{A}. In fact, although the case of non-switched systems is considered in \cite{Kong_P1}, \cite{Kong}, the idea of constructing average predictor-based control laws for systems with stochastic (switched) input delays, introduced and analyzed utilizing the backstepping method in \cite{Kong}, has inspired our developments.

In this paper, a switched linear system is considered with a constant delay in the input, where the switching signal is time dependent and almost arbitrary, as it may feature a positive, yet arbitrary dwell time. Because the future values of the switching signal may be unknown, at current time, an exact, predictor-feedback control design may be inapplicable. For this reason in this work we develop an average predictor-feedback control law. The key idea in our design is to consider an expected (average) system, with constant parameters, and compute the exact predictor of this system that the controller uses. With this approach we, essentially, derive an approximation of the exact predictor state. For this reason, in the stability analysis that relies on the backstepping, an error term appears due to the mismatch between the exact and the average predictor state. Thus, the main challenge in ensuring stability, in the presence of the mismatch in the predictors, is to derive estimates on the solutions' between the two predictor states, which we achieve in a constructive manner. In particular, we show that this mismatch is upper bounded by the norm of the (infinite-dimensional) system considered, with a bound that can be properly restricted by restricting the (maximum) difference between the original plant's parameters and the parameters of the expected system. We establish (global uniform exponential) stability of the closed-loop system, providing explicit conditions on the system's parameters, constructing a Lyapunov functional. In fact, our stability result could be alternatively viewed as a robustness result of predictor feedbacks to  arbitrary/unknown switchings of the plant parameters, in correspondence with \cite{Kong_P1} for switchings of the delay. Numerical examples are provided to illustrate the effectiveness of the controller including comparisons with nominal predictor-based controllers that may assume that the system remains in a single mode (so that a predictor-based controller can be implemented). 

The outline of the paper is as follows. Section \ref{sec2} presents the switched linear system to be examined and defines the control design. In Section \ref{sec3}, we state and prove our main result, which is exponential stability under the proposed control law. In Section \ref{sec4} we present numerical examples and in Section \ref{sec5} we provide concluding remarks.

\section{Problem Formulation and Control Design} \label{sec2}
\subsection{Switched Linear Systems with Input Delay}
We consider the following linear switched system with a constant delay in the input
\begin{equation}\label{1.1}
\dot{X}(t) = A_{\sigma(t)} X(t) + B_{\sigma(t)} U(t-D), 
\end{equation}
where $X\in \mathbb{R}^q$ is the state, $U \in \mathbb{R}$ is the control input, and $D>0$ is arbitrary long delay. We denote $L = \{0,1,2...,l\}$, as a finite index set and $\sigma:[0,+\infty)  \rightarrow L$, which is a right-continuous piecewise constant function that describes the switching signal (thus only finitely many switches can occur in any finite interval). We assume that no jump occurs in the state at a switching time and that the switching signal features arbitrary dwell time $\tau_d>0$. In this problem, at a future time $t+s$, where $s>0$, the switching rule $\sigma(t+s)$ is unknown to the user at time $t$.  
\subsection{Average Predictor-Based Control Design}
The system (\ref{1.1}) operates under the following proposed controller that we design
\begin{equation}\label{1.5}
U(t) = \bar{K} \left(e^{\bar{A}D} X(t) + \int_{t-D}^{t} {e^{\bar{A}(t-\theta)}\bar{B}U(\theta) \, d\theta}\right),
\end{equation}
where $\bar{A}$, $\bar{B}$ are viewed as some expected values of matrices $\left\{A_0,A_1,\ldots,A_l\right\}$ and $\left\{B_0,B_1,\ldots,B_l\right\} $, respectively, and can be chosen by the designer, and $\bar{K}$ is a closed-loop feedback gain, designed by the user. 
    The choice of $\bar{A}$ and $\bar{B}$ is crucial and it is dictated by the necessary restriction of the norms $|A_i - \bar{A}|, |B_i - \bar{B} |$ (see Section \ref{sec3_1} for a detailed discussion). In particular, one could, for example, employ an elaborate optimization approach, which aims at minimization of these quantities, for choosing $\bar{A}$ and $\bar{B}$. Since no a priori knowledge on the switching signal is used (because the control law (\ref{1.5}) does not employ any information about the switching rule), a computationally effective and reasonable approach for the selection of $\bar{A},\bar{B}$ would be the mean matrices of the sets $\{A_0,A_1,\ldots,A_l\}$ and $\{B_0,B_1,\ldots,B_l\}$, respectively, where as mean matrix we define the element-wise mean.
    
    One can view the choice of control law (\ref{1.5}) as corresponding to a predictor-feedback law for an expected system of the actual one, given as 
\begin{equation}\label{1.2}
\dot{{X}}(t) = \bar{A} {X}(t) + \bar{B} {U}(t-D).
\end{equation}
Solving the ODE (\ref{1.2}) for the future state, then the expected prediction of the state can be defined as
\begin{equation}\label{1.3}
\bar{P}(t) = e^{\bar{A}D} {X}(t) + \int_{t-D}^{t} {e^{\bar{A}(t-\theta)}\bar{B}{U}(\theta
) \, d\theta},
\end{equation}
and the controller is designed for the actual system (\ref{1.1}) as
\begin{equation}\label{1.4}
U(t) = \bar{K} \bar{P}(t).
\end{equation}
     The expected delay-free system (\ref{1.2}) can be stabilized with ($\ref{1.5}$) for $D=0$ under a gain $\bar{K}$ when $\bar{A}+\bar{B}\bar{K}$ can be made Hurwitz. The choice of such a gain is explained as follows. When $D=0$ the control gain is independent of the switching signal, which may lead to restrictive conditions on the matrices $A_i$, $B_i$ to guarantee closed-loop stability. However, since we consider $D$ to be potentially long, there is no clear way to choose in the predictor-based control law (\ref{1.5}) a control gain that would potentially depend on the switching signal. The reason is that due to the presence of input delay, even when (\ref{1.5}) corresponds to an exact predictor state, the choice of control gain may result, for mode $i$, in a nominal, closed-loop system in which matrix $A_i+B_iK_j$ may not be Hurwitz, as there is a mismatch between $K_j$ and a gain $K_i$ that makes $A_i+B_iK_i$ Hurwitz, where $i \neq j$, $i,j \in L$. A similar problem is also reported in, \cite{Kp}, where the authors prove stability of the switched system, under certain conditions, which involve restrictions on the upper bounds of delay values and lower bounds of the average dwell times of the system. In the presence of input delay such problems arise typically and they originate in lack of synchronization between the active mode of the system and the gain applied see, e.g., \cite{A1}. Thus, we choose a control gain $\bar{K}$ that is independent of the switching rule.

\section{Stability under average predictor-feedback }\label{sec3}
\subsection{Main Result}\label{sec3_1}
\begin{theorem}\label{Exponential Stability}
Consider the closed-loop system (\ref{1.1}) with the controller (\ref{1.5}), in which the pair $\left(\bar{A},\bar{B}\right)$ is controllable and choose $\bar{K}$ such that $\bar{A} + \bar{B}\bar{K}$ is Hurwitz. There exists $\epsilon^* > 0$ such that for any $\epsilon < \epsilon^*$, where
{\begin{equation}\label{eps} 
    \epsilon= \max \limits_{i=0, \ldots, l} \{ |A_i-\bar{A}|, |B_i-\bar{B}|\},
\end{equation}}the closed-loop system is exponentially stable in the sense that there exist positive constants $\rho$ and $\xi$ such that 
\begin{align}\label{stability equation}
\left| X(t) \right| + \sqrt{\int_{t-D}^{t} U(\theta)^2 d\theta} &\leq \rho \left( \left| X(0) \right| + \sqrt{\int_{-D}^{0} U(\theta)^2 d\theta} \right) \notag \\& \qquad \times e^{-\xi t}, \quad t \geq 0. 
\end{align}

\end{theorem}
\begin{remark}
  Theorem \ref{Exponential Stability} does not impose a restriction on the delay value or the dwell time. However, the distance between any two different matrices in the sets $\{A_0,A_1,\ldots,A_l\}$ and $\{B_0,B_1,\ldots,B_l\}$ has to be sufficiently small (see (\ref{eps_condition_2}) for an estimate of $\epsilon^*$), also depending on the delay value and the value of the dwell time. This assumption is required for two main reasons, which are related to the choice of $\bar{K}$ and of $\left( \bar{A},\bar{B} \right)$ in (\ref{1.5}). The first is due to the mismatch between the pair $\left(A_i,B_i\right)$, for $i=0,1,\ldots,l$, of the actual future mode at which the system operates and the average pair $\left(\bar{A},\bar{B}\right)$. Such a mismatch cannot be avoided by any predictor-based controller as, for arbitrary dwell time, there is no information available at current time $t$ of the future mode of the system. This also results in the requirement to restricting the maximum of $|A_i-\bar{A}|, |B_i-\bar {B}|$ as the possibility of the system operating always in a single mode, which is not known a priori, may not be excluded. Such a condition that all $\left|A_i - \bar{A} \right|$ and $\left|B_i - \bar{B} \right|$ are small, is analogous to the conditions in \cite{Kong} for the case of switching in delay values rather than in plant parameters (see also, e.g., \cite{Mazenc}, where similar conditions are employed in certain cases). The second is due to the choice of a fixed, average gain irrespectively of the different system's modes. The requirement that $\left|A_i - \bar{A} \right|$ and $\left|B_i - \bar{B} \right|$ are small guarantees the existence of a common, quadratic Lyapunov function, under the same $\bar{K}$ and under the controllability assumption on $\left(\bar{A},\bar{B}\right)$. As explained above the statement of Theorem \ref{Exponential Stability} this choice cannot be avoided either as, due to the presence of input delay, any switching signal-dependent choice of a gain could lead to a potentially closed-loop system with delay-free, nominal dynamics dictated by a non-Hurwitz matrix of the form $A_i+B_iK_j$, where $i \neq j$, $i,j \in L$. The proof of Theorem \ref{Exponential Stability}  shows that there exists a trade-off between the allowable distance among the system's matrices and the delay length, as well as the dwell time.
\end{remark}
\subsection{Proof of Theorem \ref{Exponential Stability}}
The proof of Theorem \ref{Exponential Stability} relies on some lemmas, which are presented next, together with their proofs.
\begin{lemma}\label{lemma exact predic}(Exact predictor construction.)
Let system (\ref{1.1}) experience $k$-switches within the interval $[t,t+D)$, $k \in \mathbb{N}_0$. Then for the exact predictor $P(t)$ of this system, it holds that  
\begin{align}\label{P(t)}
    P(t) &= \prod_{n=1}^{k+1} e^{A_{m_n}(s_n-s_{n-1})}X(t) + \sum_{n=1}^{k+1} \left( \prod_{j=n}^{k} e^{A_{m_{j+1}}(s_{j+1}-s_{j})}  \right. \notag \\
         &\qquad \left. \times \int_{t-D+s_{n-1}}^{t-D+s_n} e^{A_{m_n}(t-D+s_n-\theta)} B_{m_n} U(\theta) d\theta \right),
\end{align}
where $m_i \in L$, for $i=1,2,\ldots,k+1$, denotes the mode of the system before the $i$-th switching, and $s_i \in \mathbb{R}$, for $i=1,2,\ldots,k$, denotes the $i$-th switching instant, with $s_0=0$ and $s_{k+1}=D$.
\begin{proof}
    As shown in Fig. \ref{fig1}, setting $t=\theta + D$ and $P(\theta)=X(\theta+D)$, then due to this change of variables the system (\ref{1.1}) becomes   
    \begin{equation}\label{dP_theta}
        \frac{dP(\theta)}{d\theta}=A_{\sigma_{(\theta+D)}} P(\theta) + B_{\sigma_{(\theta+D)}} U(\theta).
    \end{equation}
    We divide the interval $[t, t+D]$ in intervals of constant modes, such that the system operates in $m_i$ mode if
    \begin{equation}\label{2.5}
       t-D+s_{i-1} \leq \theta \leq t-D+s_{i}, \quad i=1,2,\ldots,k+1.
    \end{equation}
    In each sub-interval the system does not exhibit switching, and thus, we set ${m_i}=\sigma(\theta+D),$  for $i=1,...,k+1$. We can now proceed to the solution of (\ref{dP_theta}) for each subsystem that is extracted from the standard form of the general solution for a time-invariant ODE system as 
    \begin{align}\label{P_theta}
        P(\theta) &= e^{{A_{m_i}}(\theta-t+D-s_{i-1})} X(t+s_{i-1}) \notag \\
        &\quad + \int_{t-D+s_{i-1}}^{\theta} e^{{A_{m_i}}(\theta - s)} {B_{m_i}}U(s) d s.
    \end{align}
    Setting $i=k+1$ in (\ref{2.5}) and $\theta=t$, from (\ref{P_theta}) we reach 
    \begin{align}\label{Pt_small}
        P(t)&=e^{A_{m_{k+1}}(D-s_{k})} X(t+s_{k}) \notag \\ &\quad +\int_{t-D-s_k}^{t} e^{A_{m_{k+1}}(t-\theta)} B_{m_{k+1}} U(\theta) d \theta . 
    \end{align}     
\begin{figure}[t]
    \centering
    \includegraphics[width=8 cm]{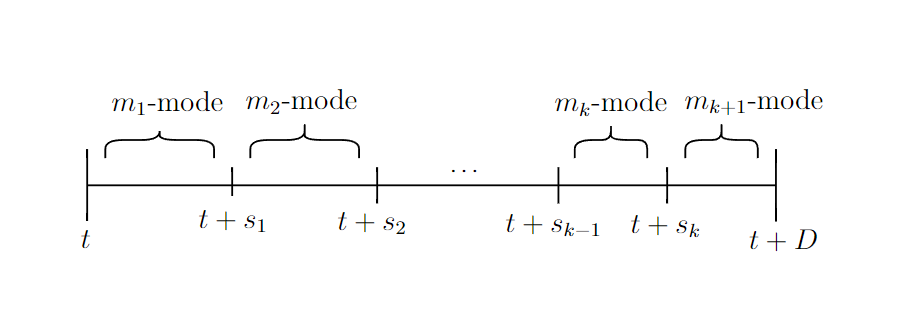}
    \caption{Switching instants and respective modes in interval $[t, t+D]$.\label{fig1}}
\end{figure}Additionally, for $\theta = t-D+s_i$ from (\ref{P_theta}), and written in terms of $X(t)$ we arrive at
    \begin{align} \label{X(t+si)}
        X(t+s_i) &= \prod_{n=1}^{i}{e^{{A_{m_n}}(s_n-s_{n-1})}} X(t) \notag \\
                 & \quad + \sum_{n=1}^{i}\prod_{j=n}^{i-1}{e^{{A_{m_{j+1}}}(s_{j+1}-s_{j})}} \notag \\ &\qquad \times \int_{t-D+s_{n-1}}^{t-D+s_n}e^{A_{m_n}(t-D+s_n-\theta)} B_{m_n}U(\theta)d\theta,
    \end{align}
    for $i=1,...,k+1$. Setting $i=k+1$, we have $X(t+s_{k+1})=X(t+D)=P(t)$. Hence from (\ref{X(t+si)}) we get (\ref{P(t)}).    
\end{proof}
\end{lemma}

\begin{lemma}(Backstepping transformation.)
The following backstepping transformation, 
    \begin{equation}\label{W_theta}
        W(\theta)=U(\theta) - \bar{K}P(\theta), \quad t-D \leq \theta \leq t,
    \end{equation}
    where $P(\theta)$ is obtained from (\ref{P_theta}) for each $\theta$ interval defined in (\ref{2.5}),  together with the control law (\ref{1.5}), transforms system (\ref{1.1}) to the target system 
    \begin{align}
        \dot{X}(t) &= \left(A_{\sigma(t)}+B_{\sigma(t)}\bar{K}\right) X(t) + B_{\sigma(t)} W(t-D) \label{Xd_trans} , \\
              W(t) &= \bar{K} \left(\bar{P}(t)-P(t)\right),\quad t\geq0, \label{W_trans}
    \end{align}
    where $\bar{P}$ and $P$ are given in (\ref{1.3}) and (\ref{P(t)}), respectively. \begin{proof}
        System (\ref{1.1}) can be written as
        \begin{align}
            \dot{X}(t) &= \left(A_{\sigma(t)} + B_{\sigma(t)}\bar{K}\right) X(t)  \notag \\ &
            \quad + B_{\sigma(t)} \left(U(t-D) -  \bar{K}X(t)\right). \label{2.16}
        \end{align}
        We adopt now (\ref{W_theta}) transformation. Setting $\theta=t-D$, from (\ref{P_theta}) we get $P(t-D)=X(t)$. Observing (\ref{P_theta}) and (\ref{2.16}), transformation (\ref{W_theta}) maps the closed-loop system consisting of the plant (\ref{1.1}) and the control law (\ref{1.5}), for all $t\geq0$, to the target system (\ref{Xd_trans}), (\ref{W_trans}).
    \end{proof}
\end{lemma}

\begin{lemma}\label{inverse_transformation}(Inverse backstepping transformation.)
The inverse backstepping transformation of $W$ is
\begin{equation}\label{inverse_theta}
    U(\theta)=W(\theta)+\bar{K}\Pi(\theta),
\end{equation}
where for $t-D \leq \theta \leq t$ divided in the sub-intervals defined in (\ref{2.5})
\begin{align}\label{2.19}
    \Pi(\theta) &= e^{({A_{m_i}+B_{m_i}\bar{K})}(\theta-t+D-s_{i-1})} X(t+s_{i-1}) \notag \\
    &\quad +\int_{t-D+s_{i-1}}^{\theta} e^{({A_{m_i}+B_{m_i}\bar{K})}(\theta - s)} {B_{m_i}}W(s) d s.
\end{align}

\begin{proof}
    Firstly, we observe from (\ref{W_theta}) that $U(\theta)=W(\theta) + \bar{K}P(\theta)$. Solving the ODE (\ref{Xd_trans}) in a similar way as (\ref{dP_theta}) it can be shown that $\Pi(\theta)=X(\theta+D)$, where $\Pi(\theta)$ is given from (\ref{2.19}), and it holds that $\Pi(\theta)=P(\theta)$. 
\end{proof}
\end{lemma}

\begin{lemma}\label{lemma_w(t)}(Bound on error due to predictor mismatch.) 
Variable $W(t)$ defined in (\ref{W_trans}) satisfies the following 
\begin{equation}\label{Wt_bound}
    | W(t) | \leq \left|\bar{K}\right| \cdot \lambda(\epsilon) \left(  |X(t)| +  \int_{t-D}^{t}{|U(\theta)|d\theta} \right), \quad t \geq 0,
\end{equation}
where $\lambda : \mathbb{R}_+ \rightarrow \mathbb{R}_+$ is a class $K_{\infty}$ function and $\epsilon$ is defined in Theorem \ref{Exponential Stability}.

\begin{proof}
    For $W(t)$ as in (\ref{W_trans}) and using (\ref{P(t)}) we can write
    \begin{equation}\label{Wt}
        W(t) = \Delta_1(t) + \Delta_2(t),
    \end{equation}
    where 
    \begin{align}
        \Delta_1(t) &= \bar{K} \left(\prod_{n=1}^{k+1}{e^{{\bar{A}}(s_n-s_{n-1})}}- \prod_{n=1}^{k+1}{e^{{A_{m_n}}(s_n-s_{n-1})}}\right)X(t), \label{D1t}\\
        \Delta_2(t) &= \bar{K}\sum_{n=1}^{k+1}\left(\prod_{j=n}^{k}{e^{{\bar{A}}(s_{j+1}-s_{j})}}\int_{t-D+s_{n-1}}^{t-D+s_n}e^{\bar{A}(t-D+s_n-\theta)} \right. \notag \\ 
                    &\qquad \times \bar{B}U(\theta)d\theta  \notag \\ 
                    &\quad - \prod_{j=n}^{k}{e^{{A_{m_{j+1}}}(s_{j+1}-s_{j})}}\int_{t-D+s_{n-1}}^{t-D+s_n}e^{A_{m_n}(t-D+s_n-\theta)} \notag \\
                    &\qquad \left. \times B_{m_n}U(\theta)d\theta\right) \label{D2t}.
    \end{align}
    For any matrix $R_i,\bar{R}$, where $R$ can be $A,B$ we set
    \begin{align}
        \Delta R_{i} &= R_{i}-\bar{R}, \label{Rmn}\\
         \epsilon_{R_{_{i}}} &= \left| \Delta R_{i} \right|, \label{epsR1}  \\
         \epsilon_R &= \max \limits_{i=0,\ldots,l} \{\epsilon_{R_{i}}\}, \label{epsR}  \\
         M_R &= \max \{|\bar{R}|,|R_0|,|R_1|,\ldots,|R_l|\}. \label{M_R}
    \end{align} 
    Setting $Y_1=A_{m_i}(s_i-s_{i-1}), \, Y_2=-\Delta A_{m_i}(s_i-s_{i-1})$,
    for $m_i$, $s_i$ defined in Lemma \ref{lemma exact predic}, and using the fact that for any two $n \times n$ matrices $Y_1$, $Y_2$ the following inequality holds 
    \begin{align} \label{series}
       \left| e^{Y_1+Y_2}-e^{Y_1} \right| \leq |Y_2| e^{ |Y_1| }e^{ |Y_2| },
    \end{align} 
    where $|\cdot|$ denotes an arbitrary matrix norm\footnote{Proof of (\ref{series}) relies on the power series expansion for the matrix exponential and triangle inequality, see, e.g., \cite{Lie}.}, we have from (\ref{Rmn})--(\ref{series}) and since $(s_n-s_{n-1}) \leq D$, that
    \begin{equation}\label{TheorResult}
        \left |e^{\bar{A}(s_n-s_{n-1})} - e^{A{m_n} (s_n-s_{n-1})} \right | \leq \epsilon_{A} \cdot D \cdot e^{ M_A (s_n-s_{n-1}) } e^{ \epsilon_{A} D }.
    \end{equation}  
    We now upper bound the expression from (\ref{D1t}). We define
    \begin{equation}\label{Tk+1}
        T_{k+1}=\left| \prod_{n=1}^{k+1}{e^{{\bar{A}}(s_n-s_{n-1})}}- \prod_{n=1}^{k+1}{e^{{A_{m_n}}(s_n-s_{n-1})}}  \right|.
    \end{equation}
    Developing (\ref{Tk+1}) for each iteration, for $k=0$, the result in (\ref{TheorResult}) can be directly applied to (\ref{Tk+1}). For $k=1$,
        \begin{align}
            T_2 = \left| e^{\bar{A}s_1}e^{\bar{A}(s_2-s_1)} - e^{A{m_1}s_1} e^{A{m_2}(s_2-s_1)} \right|.
        \end{align}
        We expand the difference within the norm and using the the triangle inequality for the norm bounds we get \begin{align}\label{n=2}
            T_2 & \leq \left| e^{\bar{A}s_1} \right| \cdot \left| e^{\bar{A}(s_2-s_1)} - e^{A{m_2} (s_2-s_1)} \right|  + \left| e^{A{m_2} (s_2-s_1)} \right| T_1 .
        \end{align}
        We apply now (\ref{TheorResult}) to (\ref{n=2}), to obtain
        \begin{align}
            T_2 \leq \epsilon_{A} \cdot 2D \cdot e^{ M_A s_2 } e^{ \epsilon_{A} D } .
        \end{align}
        For some $k$, we assume that the following expression holds
        \begin{align}\label{tkResult}
            T_{k} \leq \epsilon_{A} \cdot kD \cdot e^{ M_A s_k } e^{ \epsilon_{A} D } .
        \end{align}
    We prove that the formula holds generally using the induction method.     
    Since (\ref{n=2}), (\ref{tkResult}) hold, expanding (\ref{Tk+1}) and using the triangle inequality we get
    \begin{align}\label{tk+1last}
        T_{k+1} & \le \left| e^{\bar{A}(s_{k+1}-s_k)}-e^{A_{m_{k+1}}(s_{k+1}-s_k)}\right|\left|\prod_{n=1}^{k}{e^{{\bar{A}}(s_n-s_{n-1})}}\right| \notag \\ 
               &\quad + \left|e^{A_{m_{k+1}}(s_{k+1}-s_k)}\right| T_k. 
    \end{align}
    Applying (\ref{TheorResult}) and (\ref{tkResult}) to (\ref{tk+1last}) we get
    \begin{align}\label{tk+1Result}
        T_{k+1} & \leq \epsilon_{A} \cdot (k+1) D \cdot e^{ M_A \cdot (s_{k+1}) } e^{ \epsilon_{A} D },
    \end{align}
    which makes (\ref{tkResult}) legitimate for all $k$.
    Hence, applying any arbitrary matrix norm and substituting (\ref{tk+1Result}) in (\ref{D1t})  we get 
    \begin{equation}\label{D1_Bound}
        |\Delta_1(t)| \leq \delta_1 |\bar{K}| |X(t)|,
    \end{equation}
    where \begin{equation}
        \delta_1 = \epsilon \cdot \left(\left\lceil \frac{D}{\tau_d} \right\rceil+1\right) D \cdot e^{D \left( M_A + \epsilon \right) },
    \end{equation}
    since
    \begin{equation}\label{switching bound}
        k \leq   \left\lceil \frac{D}{\tau_d} \right\rceil.
    \end{equation}For the (\ref{D2t}) expression we start by applying the property $M'N'-MN=M'(N'- N) + (M'-M)N$ in each part where $M,N$ are arbitrary matrices, and hence, (\ref{D2t}) becomes 
    \begin{align}\label{d2t_new}
        \Delta_2(t) &= \bar{K} \sum_{n=1}^{k+1}\left\{ \prod_{j=n}^{k} { e^{{\bar{A}}(s_{j+1}-s_{j})}}\left(Z_{1,n}(t) +Z_{2,n}(t)\right)  \right. \notag \\ & \left. \quad + Z_{3,n}(t)\right\}, 
    \end{align}
    where
    \begin{align}
        Z_{1,n}(t) &=  \int_{t-D+s_{n-1}}^{t-D+s_n}{e^{\bar{A}(t-D+s_n-\theta)}(\bar{B}-B_{m_n})U(\theta)d\theta},\label{Z1}  \\
        Z_{2,n}(t) &=  \int_{t-D+s_{n-1}}^{t-D+s_n}\left( e^{\bar{A}(t-D+s_n-\theta)} -e^{A_{m_n}(t-D+s_n-\theta)} \right) \notag \\ 
        &\qquad \times B_{m_n}U(\theta)d\theta, \label{Z2}\\
        Z_{3,n}(t) &=  {\left(\prod_{j=n}^{k}{e^{{\bar{A}(s_{j+1}-s_{j})}}}-\prod_{j=n}^{k}{e^{{A_{m_{j+1}}}(s_{j+1}-s_{j})}}\right)} \notag \\ 
        &\qquad \times \int_{t-D+s_{n-1}}^{t-D+s_n}e^{A_{m_n}(t-D+s_n-\theta)}B_{m_n}U(\theta)d\theta  .  \label{Z3}
    \end{align}
    We proceed with bounding separately the above terms. Applying (\ref{epsR1}) and any norm in (\ref{Z1}) we get
        \begin{align}\label{Z1bnd}
            \left| Z_{1,n}(t) \right| \leq e^{ |\bar{A}|(s_n-s_{n-1})} \cdot  \epsilon_{B_{m_n}} \cdot \int_{t-D+s_{n-1}}^{t-D+s_n}{|U(\theta)|d\theta}.
        \end{align}
        Similarly for (\ref{Z2}) we recall (\ref{epsR1})--(\ref{M_R}), (\ref{TheorResult}), and hence,
        \begin{align}\label{Z2bnd}
            \left| Z_{2,n}(t) \right| &\leq \epsilon_A \cdot D \cdot  e^{M_A (s_n-s_{n-1})} e^{\epsilon_A \cdot D} M_B \notag \\ & \quad  \times \int_{t-D+s_{n-1}}^{t-D+s_n}{|U(\theta)|d\theta}.
        \end{align}
        Recalling (\ref{tk+1Result}) then, similarly for (\ref{Z3}), we get 
        \begin{align}\label{Z3bnd}
            {\left| Z_{3,n}(t) \right|} &\leq \epsilon_A \cdot e^{M_A(s_{k+1-n})} \cdot (k+1-n)D \cdot e^{\epsilon_A \cdot D} \notag \\
            &\quad \times e^{M_A \cdot (s_n-s_{n-1})} M_B \int_{t-D+s_{n-1}}^{t-D+s_n}{|U(\theta)|d\theta}.
        \end{align}
    Applying (\ref{switching bound}), (\ref{Z1bnd})--(\ref{Z3bnd}) in (\ref{d2t_new}) we get 
    \begin{equation}\label{D2_Bound}
       \left| \Delta_2(t) \right| \leq |\bar{K}| \cdot \delta_2 \int_{t-D}^{t}{|U(\theta)|d\theta},
    \end{equation}
    where
    \begin{align}\label{deltab}
        \delta_2 &= \epsilon \cdot e^{M_A D} \left( e^{\epsilon D} D M_B \left[1+ e^{M_A D} \left(   \left\lceil \frac{D}{\tau_d} \right\rceil +1\right) \right] +1\right) . 
    \end{align}
    Applying (\ref{D1_Bound}), (\ref{D2_Bound}) in (\ref{Wt}) we arrive at (\ref{Wt_bound}) where
    \begin{align}
        \lambda(\epsilon) &= \max \{\delta_1(\epsilon),\delta_2(\epsilon)\}.
    \end{align}
\end{proof}
\end{lemma}

\begin{lemma}\label{lemma_u(theta)}(Norm equivalency.)
    For the inverse transformation (\ref{inverse_theta}), (\ref{2.19}) the following inequality holds for some positive constant $\nu_1$ 
    \begin{equation}\label{ut_bound}
       \int_{t-D}^{t}{| U(\theta) |^2 d \theta} \leq \nu_1 \left( |X(t)|^2 + \int_{t-D}^{t}{| W(\theta) |^2 d \theta} \right).
    \end{equation}
    Similarly, for the direct transformation (\ref{P_theta}), (\ref{W_theta}) it holds for some positive constant $\nu_2$
    \begin{equation}\label{wt_bound}
       \int_{t-D}^{t}{| W(\theta) |^2 d \theta} \leq \nu_2 \left( |X(t)|^2 + \int_{t-D}^{t}{| U(\theta) |^2 d \theta} \right).
    \end{equation}
\begin{proof}
    From (\ref{inverse_theta}) for the inverse transformation we apply Young's inequality to obtain 
    \begin{align}\label{U_theta_bound}
        \int_{t-D}^{t}{| U(\theta) |^2 d \theta} &\leq 2 \left( \int_{t-D}^{t}{ |W(\theta) |^2 d\theta} \right. \notag \\ 
        &\left. \quad + |\bar{K}|^2  \int_{t-D}^{t}{  \left| { \Pi(\theta)   } \right| ^2 d \theta}  \right).
    \end{align}
    Setting $H_i=A_i + B_i \bar{K}$, for $i=0,\ldots,l$, and using (\ref{2.5}), (\ref{2.19}) and (\ref{M_R}), we obtain the following,
    \begin{align}\label{T_t}
        &\int_{t-D}^{t}{| \Pi(\theta) |^2 d \theta} = \sum_{i=1}^{k+1} \int_{t-D+s_{i-1}}^{t-D+s_i} \left| e^{H_{m_i}(\theta -t+D-s_{i-1})} \right. \notag \\ 
        & \quad \times \left. \left( \prod_{n=1}^{i-1} e^{H_{m_n} (s_n - s_{n-1})} X(t) + \sum_{n=1}^{i-1}\prod_{j=n}^{i-2}{e^{{H_{m_{j+1}}}(s_{j+1}-s_{j})}} \right. \right.  \notag \\
        & \quad \left. \left.  \times \int_{t-D+s_{n-1}}^{t-D+s_n}e^{H_{m_n}(t-D+s_n-\theta)}B_{m_n}W(\theta)d\theta \right) \right. \notag \\
        & \quad \left. + \int_{t-D+s_{i-1}}^{\theta}e^{H_{m_i}(\theta - s)}B_{m_i}W(s)ds\right|^2 d\theta.    
    \end{align}
     Applying the triangle inequality in (\ref{T_t}) and substituting in (\ref{U_theta_bound}) we reach (\ref{ut_bound}), where
    \begin{align}
        \nu_1 &= 2 \max \left\{ 2 \bar{K}^2 D e^{2 M_H D}, 1+2 \bar{K}^2  D^2 e^{2 M_H D} M_{B}^2 \right\}.
    \end{align}    
    Analogously, using the direct transformation from (\ref{P_theta}) and (\ref{W_theta}), we can similarly prove (\ref{wt_bound}) via (\ref{P(t)}), where
    \begin{align}
        \nu_2 &= 2 \max \left\{ 2 \bar{K}^2 D e^{2 M_A D}, 1+2 \bar{K}^2 D^2 e^{2 M_A D} M_{B}^2 \right\}.
    \end{align}\
\end{proof}
\end{lemma}

\begin{lemma}\label{trans_stability}(Stability of target system.)
    Let the pair $\left(\bar{A},\bar{B}\right)$ be controllable and choose $\bar{K}$ such that $\bar{A} + \bar{B}\bar{K}$ is Hurwitz, and thus, it holds that   
    \begin{equation}\label{mean_delayf_stability}
        \left(\bar{A} + \bar{B} \bar{K}\right)^T P + P \left(\bar{A} + \bar{B} \bar{K}\right) = -Q,
    \end{equation}
    for some \( P \)=\( P^T > 0\), \( Q \)=\( Q^T > 0\). For any $\epsilon < \epsilon^*$, where 
    \begin{align}\label{eps_condition_2}
        \epsilon^* &= \min \left\{ \frac{ \lambda_{\min}(Q)  }{2 |P| \left(1+|\bar{K}|\right)} , \alpha^{-1} 
        \left( \frac{\lambda_{\min}(Q)}{2}  \right), \notag \right. \\
        &\quad \left. \lambda^{-1}\left(\frac{1}{|\bar{K}| \sqrt{2 e^D D \nu_1}} \right)  \right\},
    \end{align}  
    for $\lambda(\epsilon)$ defined in Lemma \ref{lemma_w(t)}, and $\alpha$ be class $K$ function on $\left[0,\frac{\lambda_{\min}({Q})}{2 |P| \left(1+|\bar{K}|\right)}\right)$ defined as 
    \begin{equation}\label{a function}
        \alpha(\epsilon)= \epsilon |P|\left(1+|\bar{K}|\right) + \lambda^2(\epsilon) \frac{ 4\left( \left|P\right| M_B \right)^2 e^D |\bar{K}|^2 (D \nu_1 + 1) }{ \lambda_{\min}(Q) - 2 \epsilon |P|\left(1+|\bar{K}|\right)  },    \end{equation}
    the target system (\ref{Xd_trans}), (\ref{W_trans}), is exponentially stable, in the sense that there exist positive constants $\kappa$ and $\mu$ such that
    \begin{align}\label{trans_stability_result}
        \left| X(t) \right|^2 + \int_{t-D}^{t} W(\theta)^2 d\theta &\leq \kappa \left( |X(0)|^2 \right. \notag \\ 
        & \left. \quad  + \int_{-D}^{0} W(\theta)^2 d\theta \right) e^{-\mu t}, \ t \geq 0.
    \end{align}

\begin{proof}
    For the target system we can now adopt the following Lyapunov functional
    \begin{equation}\label{LyapunovFnc}
        V(t) = X(t)^T P X(t) + b \int_{t-D}^{t} e^{(\theta+D-t)}W(\theta)^2 d\theta.
    \end{equation}
    Recall that, for the switched system (\ref{1.1}) using (\ref{eps}), (\ref{Rmn}), and (\ref{mean_delayf_stability}), for any vector $y \neq 0$ it holds that 
    \begin{align}\label{lyap_exact_sys}
        y^T & \left[ \left(A_i + B_i \bar{K}\right)^T P + P \left(A_i + B_i \bar{K}\right) \right] y \leq \notag \\ 
        & \qquad - y^T \left[  Q - 2 \epsilon |P|\left(1+|\bar{K}|\right) {I} \right] y .
    \end{align} 
    The matrix $Q - 2 \epsilon |P|\left(1+|\bar{K}|\right) {I} $ is positive definite when 
    \begin{equation}\label{eps_assume}
        \epsilon < \frac{\lambda_{\min}({Q})}{2 |P| \left(1+|\bar{K}|\right)}.
    \end{equation}
    Calculating the derivative of (\ref{LyapunovFnc}), across the solutions of the target system, under (\ref{lyap_exact_sys}) we reach at
    \begin{align}
        & \dot{V}(t) \leq -X(t)^T \left( Q - 2 \epsilon |P|\left(1+|\bar{K}|\right) {I} \right) X(t)   \notag \\ 
        &\ + B_{\sigma(t)}^TW(t-D)^T  P X(t) + X(t)^T P B_{\sigma(t)} W(t-D)  \notag \\
        &  + b(\epsilon)e^D W(t)^2 - b W(t-D)^2 - b\int_{t-D}^{t} e^{(\theta+D-t)}W(\theta)^2 d\theta. \label{dv_t_1}
    \end{align}
    We further observe 
    \begin{align}
            -X(t)^T & \left( Q - 2 \epsilon |P|\left(1+|\bar{K}|\right) {I} \right) X(t) \leq \notag \\
            &\quad - \left( \lambda_{\min}(Q) - 2 \epsilon |P|\left(1+|\bar{K}|\right) \right)|X(t)|^2,
    \end{align}
    and
    \begin{align}\label{BTWT}
        B_{\sigma(t)}^T  W(t-D)^T & P X(t) + X(t)^T PB_{\sigma(t)} W(t-D) \leq \notag \\
        & 2 \left|X(t)^T\right| \left|P\right| M_B \left|W(t-D)\right|.
    \end{align}
    Applying Young's inequality in (\ref{BTWT}) and bounding (\ref{dv_t_1}), if we choose $b$ as
    \begin{equation}
        b(\epsilon)=\frac{ 2\left( \left|P\right| M_B \right)^2 }{ \lambda_{\min}(Q) - 2 \epsilon |P|\left(1+|\bar{K}|\right)  }, 
    \end{equation}
    then we get from (\ref{dv_t_1}) that
    \begin{align}
    \label{dv_t_11}
        \dot{V}(t) &\leq -\frac{1}{2}\left( \lambda_{\min}(Q) - 2 \epsilon |P|\left(1+|\bar{K}|\right) \right)|X(t)|^2 \notag \\
        & \quad + b(\epsilon)e^D W(t)^2 - b(\epsilon)\int_{t-D}^{t} W(\theta)^2 d\theta.
    \end{align}
    Using the result from Lemma \ref{lemma_w(t)} and applying Young's and the Cauchy-Schwarz inequalities we obtain
    \begin{align}\label{wt_sqrt1}
        W(t)^2 \leq 2 |\bar{K}|^2 \lambda^2(\epsilon)\left(|X(t)|^2 +  D \int_{t-D}^{t}{U(\theta)^2 d\theta} \right).
    \end{align}
     Applying the result from Lemma \ref{lemma_u(theta)} we get from (\ref{wt_sqrt1}) that  
    \begin{align}\label{wt_sqrt11}
        W(t)^2 &\leq 2 |\bar{K}|^2 \lambda^2(\epsilon)(D \nu_1 + 1) |X(t)|^2 + 2|\bar{K}|^2 \lambda^2(\epsilon) D \nu_1 \notag \\
        &\qquad \times \int_{t-D}^{t}{W(\theta)^2 d\theta}.
    \end{align}    
    Expanding (\ref{dv_t_11}) and employing (\ref{wt_sqrt11})  we get
    \begin{align}\label{dv_t_12}
        \dot{V}(t) &\leq - \left( \frac{1}{2}\left( \lambda_{\min}(Q) - 2 \epsilon |P|\left(1+|\bar{K}|\right) \right) - 2 b(\epsilon) e^D |\bar{K}|^2 \right. \notag \\
        &\qquad \left. \times \lambda^2(\epsilon)(D \nu_1 + 1) 
        \right)|X(t)|^2 \notag \\
                   &\quad -b(\epsilon) \int_{t-D}^{t}{ \left[1 - 2 e^D |\bar{K}|^2 \lambda^2(\epsilon) D \nu_1         \right] W(\theta)^2 d\theta } . 
    \end{align}
    In order to preserve negativity in (\ref{dv_t_12}) it is required that
    \begin{align}\label{d_v_t13}
        \frac{1}{2} \lambda_{\min}(Q) -  \epsilon |P|\left(1+|\bar{K}|\right) &- 2 b(\epsilon) e^D |\bar{K}|^2\lambda^2(\epsilon) \notag \\ 
        &\qquad \times (D \nu_1 + 1) > 0  
    \end{align}
    and
    \begin{equation}\label{d_v_t14}
        1 - 2 e^D |\bar{K}|^2 \lambda^2(\epsilon) D \nu_1 > 0 ,
    \end{equation}
    which hold under the restriction on $\epsilon$ in the statement of the lemma.
    From (\ref{a function}) and (\ref{d_v_t13}) we get
    \begin{align}
        \dot{V}(t) &\leq - \min \left \{ 1-2e^D|\bar{K}|^2\lambda^2(\epsilon) D \nu_1, \right. \notag \\ 
        &\quad \left. \frac{ \frac{1}{2}  \lambda_{\min}(Q) - \alpha(\epsilon )}{\lambda_{\max}(P)} \right \} V(t).
    \end{align}
    Now we can apply the comparison principle, and hence,
    \begin{equation}
        V(t) \leq e^{-\mu t}V(0),\quad t \geq 0,
    \end{equation}
    where 
    \begin{align}
        \mu &= \min \left \{ 1-2e^D|\bar{K}|^2\lambda^2(\epsilon) D \nu_1, \frac{ \frac{1}{2}  \lambda_{\min}(Q) - a(\epsilon) }{\lambda_{\max}(P)} \right \}.
    \end{align}
    Exponential stability for the system in the $(X, W)$ variables can now be proved. Observing (\ref{LyapunovFnc}) we have
    \begin{align}\label{2.100}
    \mu_1 & \left( |X(t)|^2 + \int_{t-D}^{t} W(\theta)^2 d\theta \right) \leq V(t) \notag  \\
    &\leq \mu_2 \left( |X(t)|^2 + \int_{t-D}^{t} W(\theta)^2 d\theta \right),
    \end{align}
    where
    \begin{equation}
    \begin{aligned}
    \mu_1 &= \min \left\{ \lambda_{\min}(P), \frac{2|PB|^2}{ \lambda_{\min}(Q) - 2 \epsilon |P|\left(1+|\bar{K}|\right) } \right\}, \\
    \mu_2 &= \max \left\{ \lambda_{\max}(P), \frac{2|PB|^2}{\lambda_{\min}(Q) - 2 \epsilon |P|\left(1+|\bar{K}|\right)} e^D \right\}.
    \end{aligned}
    \end{equation}  
    Setting $\kappa = \frac{\mu_2}{\mu_1}$ we reach (\ref{trans_stability_result}).
\end{proof}
\begin{customproof}[{Proof of Theorem \ref{Exponential Stability}}. ]
 Now we are able to complete the proof for Theorem \ref{Exponential Stability} and hence, conclude the stability of the original system. Combining (\ref{wt_bound}), (\ref{2.100}), and the result in Lemma \ref{lemma_u(theta)}, we get (\ref{stability equation}) where $\rho = \sqrt{\frac{2\mu_1\nu_1\nu_2}{\mu_2}}, \
    \xi  = \frac{\mu}{2}.$ 
\end{customproof}
\end{lemma}

\section{Numerical Simulation Results}\label{sec4}
\begin{figure}[ht]
    \centering
    \includegraphics[width=8 cm]{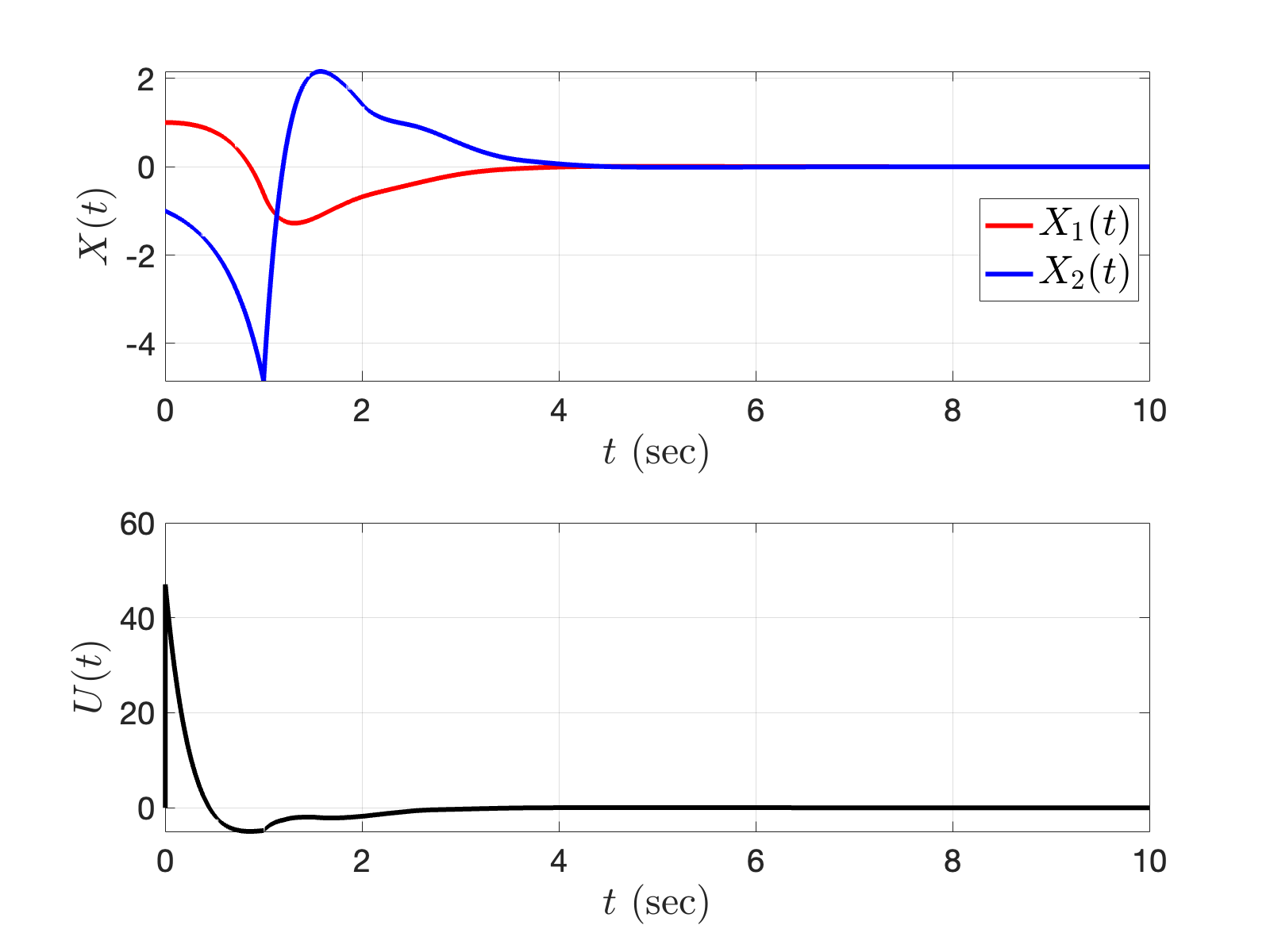}
    \caption{Evolution of state $X(t)$ and control input $U(t)$ for system (\ref{1.1}) with (\ref{ex1}), (\ref{ex1_dyn}), under controller (\ref{1.5}) with (\ref{ex1_k}). 
    \label{fig3}}
    \end{figure}

\begin{figure}[ht]
    \centering
    \includegraphics[width=8 cm]{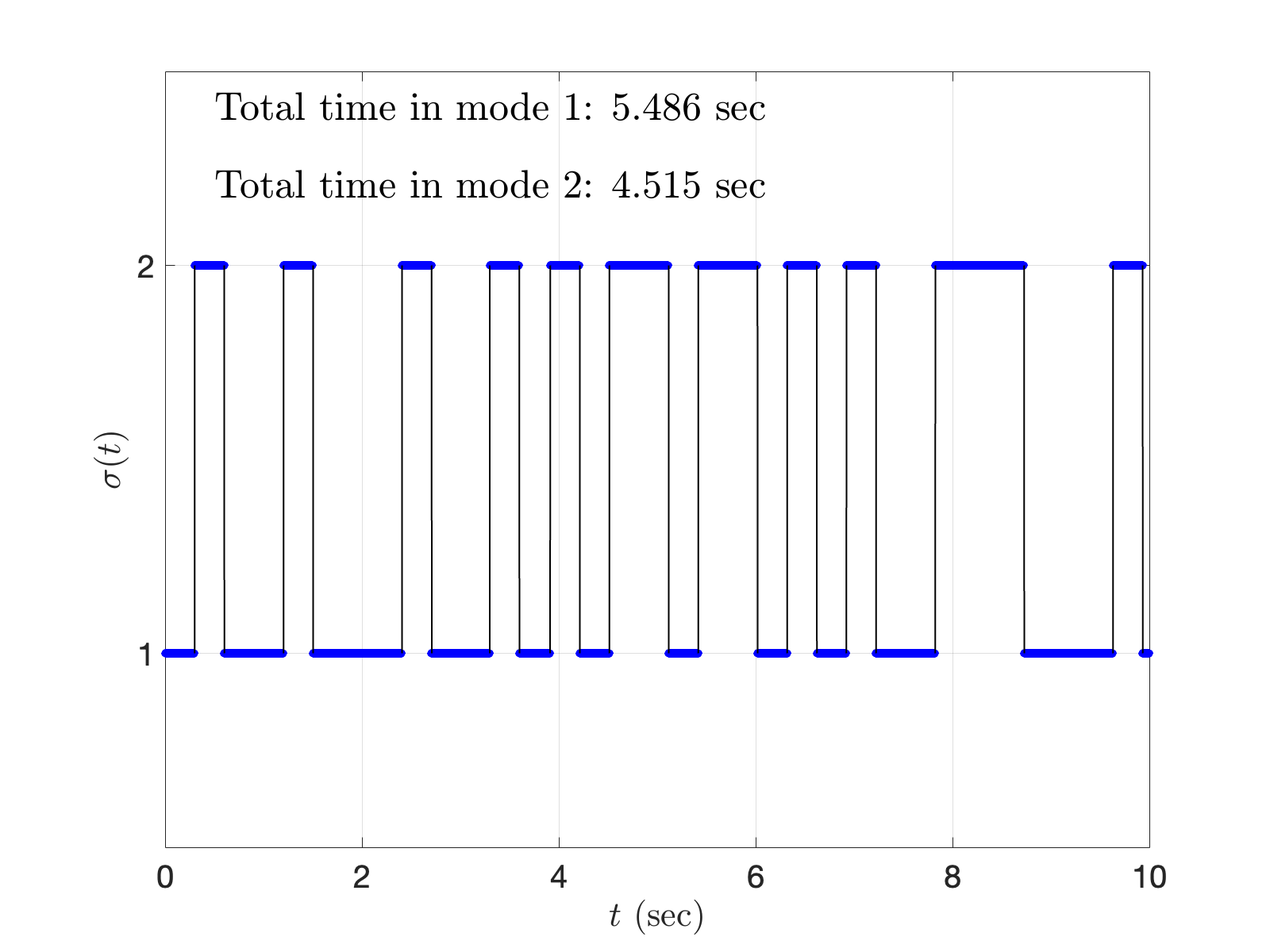}
    \caption{Evolution of switching signal $\sigma(t)$ for all the case studies.
    \label{fig2}}
\end{figure}

Consider the switched system (\ref{1.1}) with the subsystems \( A_1 \) and \( A_2 \) defined as
\begin{equation}\label{ex1}
    A_1 = \begin{bmatrix}
    1 & 1 \\
    1 & 2
\end{bmatrix}, \quad
A_2 = \begin{bmatrix}
    1.01 & 0.99 \\
    1.01 & 2.01
\end{bmatrix},
\end{equation}
and input matrices \( B_1 \) and \( B_2 \) as
\begin{equation}\label{ex1_dyn}
    B_1 = \begin{bmatrix}
    0 \\
    1
\end{bmatrix}, \quad
B_2 = \begin{bmatrix}
    0 \\
    1
\end{bmatrix}.
\end{equation}
Next we compute the expected values $\bar{A},\bar{B}$ for the controller (\ref{1.5}), resulting in $\epsilon=0.0071$, and choose
\begin{equation}\label{ex1_k}
   \bar{K}= \begin{bmatrix}
    -13.1005 & -8.01
\end{bmatrix}.
\end{equation}
This gain places the closed-loop poles of the expected system at $-3,-2$ resulting in $\epsilon^*=0.0093$. (computed using (\ref{eps_condition_2})). Setting $D=1$, $\tau_d=0.3$, and initial conditions $X_0 = \begin{bmatrix} 1 & -1 \end{bmatrix}$, $U(s)=0$, for $s \in [-D,0)$, we obtain the system's behavior depicted in Fig. \ref{fig3}. Fig. \ref{fig2} illustrates the evolution of the switching signal over the total simulation time, where the blue lines indicate the active mode and the black lines the switching instants. 

\begin{figure}[ht!]
    \centering
    \includegraphics[width=8 cm]{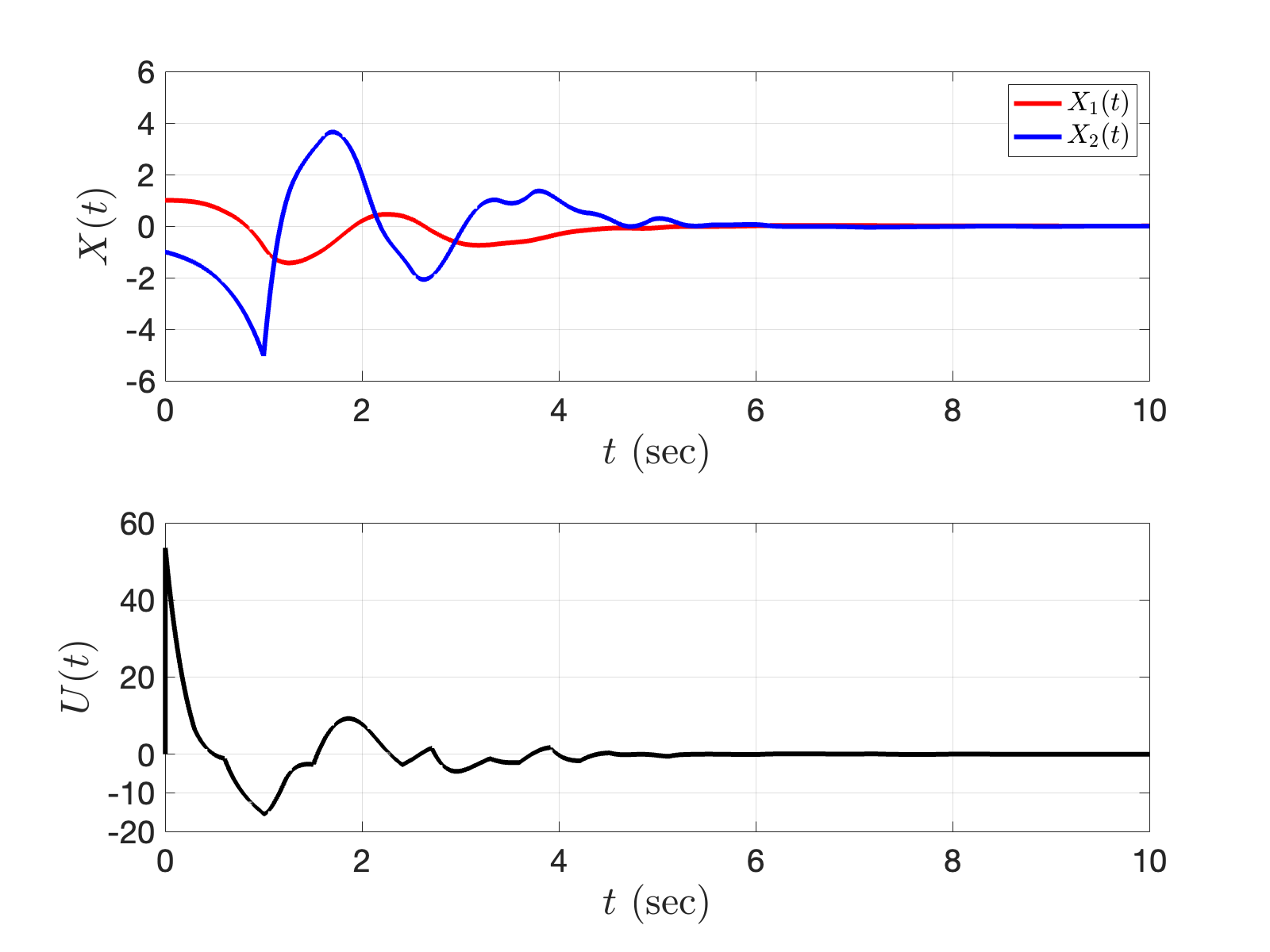}
    \caption{Evolution of state $X(t)$ and control input $U(t)$ for system (\ref{1.1}) with (\ref{ex2}), (\ref{ex1_dyn}), under controller (\ref{1.5}) with (\ref{ex2_k}). 
    \label{fig4}}
    \end{figure}

For increased discrepancy between the subsystems we consider
\begin{equation}\label{ex2}
    A_1 = \begin{bmatrix}
    1 & 1 \\
    1 & 2
\end{bmatrix}, \quad
   A_2 = \begin{bmatrix}
    1.07 & 1.15 \\
    1.06 & 2.09
\end{bmatrix},
\end{equation}
which gives (using (\ref{eps})) $\epsilon=0.0986$, and choose \begin{equation}\label{ex2_k}
    \bar{K}= \begin{bmatrix}
    -12.4218 & -8.08
\end{bmatrix},
\end{equation}
 for the same pole placement of the closed-loop, expected system. This choice results in $\epsilon^*=0.0109$, obtaining the behaviour of the system as shown in Fig. \ref{fig4}. Although the stability condition is violated, exponential convergence is maintained, but with a degraded transient response. This indicates that $\epsilon^*$ in (\ref{eps_condition_2}) is conservative. To illustrate the usefulness of the average predictor-feedback law (\ref{1.5}), we compare the behaviour of the same system, with matrices (\ref{ex2}), (\ref{ex1_dyn}), operating under the following controllers, for $i=1,2$
\begin{equation}\label{onecontrol}
U(t) = \bar{K} \left(e^{{A_i}D} X(t) + \int_{t-D}^{t} {e^{{A_i}(t-\theta)}B_iU(\theta) \, d\theta}\right).
\end{equation}
These controllers assume that the system operates always only in one mode. Thus, the test validates the effectiveness of the controller (\ref{1.5}), which accounts, in an average manner, for the different systems' dynamics.
\begin{figure}[ht!]
    \centering
    \includegraphics[width=8 cm]{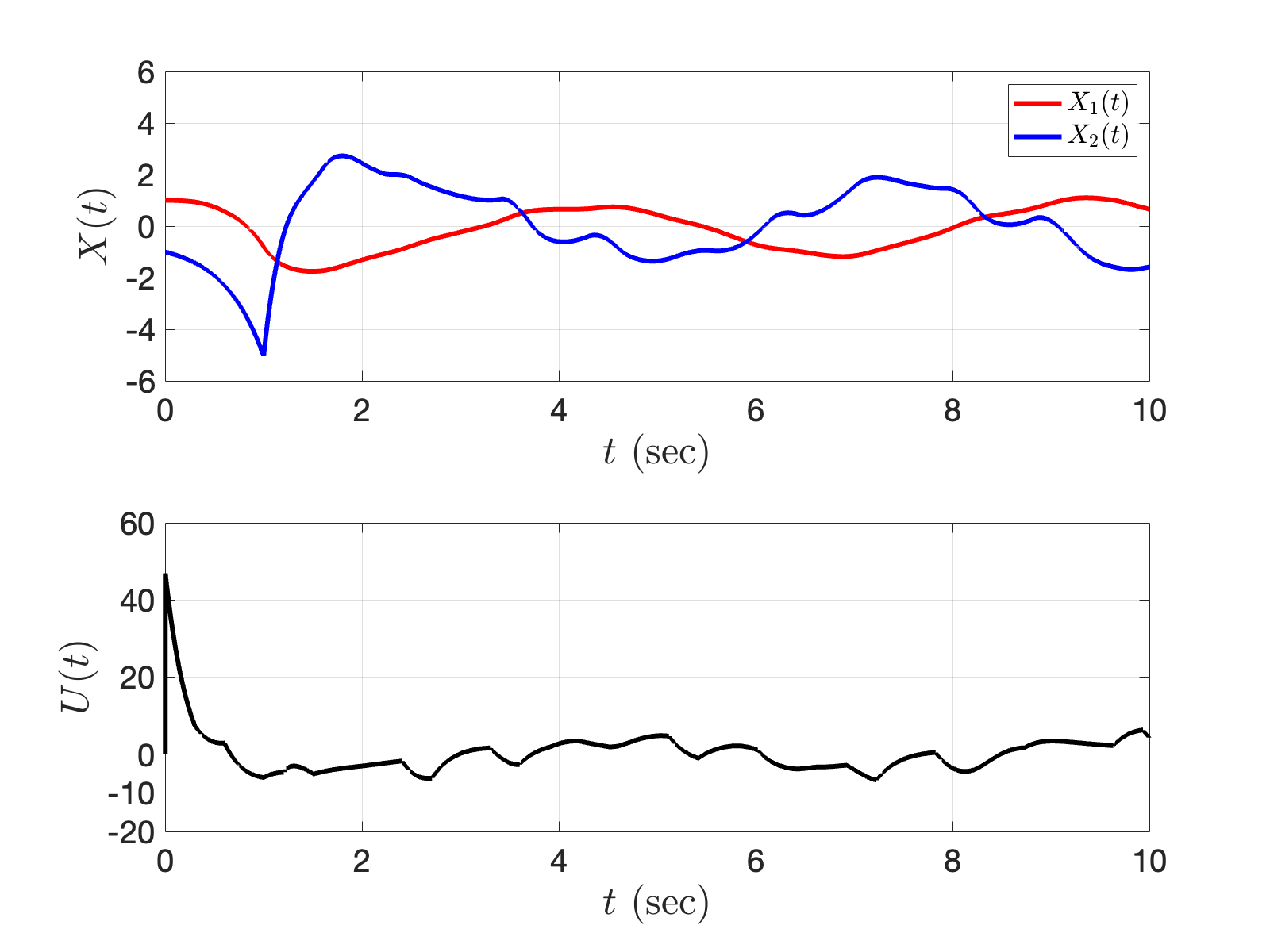}
    \caption{Evolution of state $X(t)$ and control input $U(t)$ for system (\ref{1.1}) with (\ref{ex2}), (\ref{ex1_dyn}), under controller (\ref{onecontrol}) for $i=1$ with (\ref{ex2_k}).
    \label{fig5}}
    \end{figure}
Fig. \ref{fig5} shows the behaviour of the system operating under the controller (\ref{onecontrol}) for $i=1$ and Fig. \ref{fig6} shows the behaviour of the system operating under the same controller for $i=2$. It can be seen that in both cases the performance of the respective closed-loop systems is not the desired one. Examining Fig. \ref{fig2}, we observe that the system remains in mode $1$ for a slightly longer duration. Consequently, it is expected that the controller (\ref{onecontrol}) for $i=1$, which assumes that the system remains at mode $1$, demonstrates generally, a better performance, particularly for times smaller than $5s$, where the system operates at mode $1$ for much longer time periods. This expectation is confirmed as shown in Fig. \ref{fig5}, while, as shown in Fig. \ref{fig6}, for times (roughly) larger than $5s$ the performance of the closed loop system under (\ref{onecontrol}) for $i=2$ is improved because the system starts operating at mode $2$ for longer time periods. In general, the switching dynamics still prevent the controller from effectively stabilizing the state for the case corresponding to Fig. \ref{fig5}, indicating that mode $2$ significantly influences the system’s dynamics despite its overall shorter operating time. This demonstrates the necessity to employ the average predictor-feedback law (\ref{1.2}). Note that if one chooses a different gain in (\ref{onecontrol}), for example, depending on $i$ such that the nominal matrices $A_i + B_i K_i$ have eigenvalues at the same points as $\bar{A} + \bar{B}\bar{K}$, we observed in simulation that the closed-loop system performance deteriorates. This is however also dependent on the specific scenario considered.
\begin{figure}[ht!]
    \centering
    \includegraphics[width=8 cm]{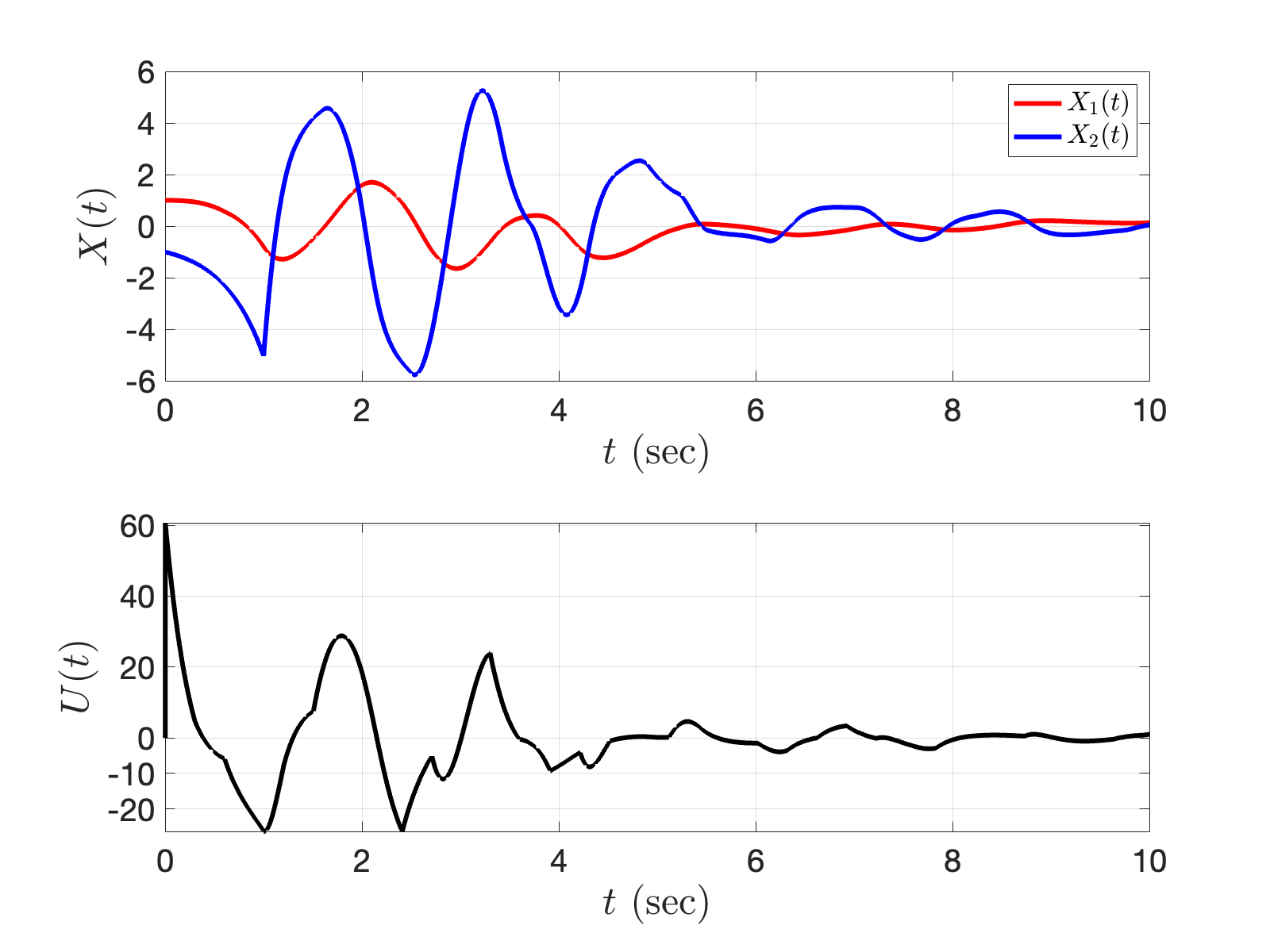}
    \caption{Evolution of state $X(t)$ and control input $U(t)$ for system (\ref{1.1}) with (\ref{ex2}), (\ref{ex1_dyn}), under controller (\ref{onecontrol}) for $i=2$ with (\ref{ex2_k}).
    \label{fig6}}
    \end{figure}
    
\section{Conclusions and Discussion}\label{sec5} 
In this work, we have developed a control design and analysis method for switched linear systems with input delays, where the switching signal is time-dependent with an arbitrary dwell time and its future values may not be available. An average, predictor-based feedback control law is introduced and conditions are derived on the plant's parameters, guaranteeing the proximity with the exact predictor-feedback law. Exponential stability is proved relying on backstepping and based on a Lyapunov functional construction. Numerical simulations confirm the effectiveness of the proposed control law. We are currently investigating alternative, average predictor-based control designs, which may impose less restrictive conditions on the plant parameters by taking into account in the average, predictor state construction time horizons of exact prediction, utilizing the potential knowledge of a dwell time and potentially restricting its value (that are not assumed here). We also investigate alternative choices for $\bar{A}$ and $\bar{B}$ utilizing optimization routines.

\bibliography{ifacconf}             

\end{document}